\documentclass[aps,onecolumn,preprint,superscriptaddress,nofootinbib]{revtex4}
\usepackage{amsmath,amssymb,color,mathrsfs, graphicx,verbatim,epsfig, bbm, wasysym, pstricks}
\newcommand{\beq}{\begin{equation}}
\newcommand{\eeq}{\end{equation}}
\newcommand{\beqa}{\begin{eqnarray}}
\newcommand{\eeqa}{\end{eqnarray}}
\newcommand{\tr}{{\rm Tr}}

\newcommand{\mpl}{M_{{\rm Pl}}}

\begin{document}
\begin{flushright}UMD-PP-09-036 \\
\end{flushright}
\vspace{0.2cm}

\title{Uplifted Metastable Vacua and Gauge Mediation in SQCD}

\author{ Amit Giveon\footnote{Permanent address: Racah Institute of Physics, The Hebrew University, Jerusalem 91904, Israel.}}
\affiliation{EFI and Department of Physics, University of Chicago,\\
5640 S. Ellis Av., Chicago, IL 60637}

\author{ Andrey Katz}
\affiliation{Department of Physics, University of Maryland,\\
College Park, MD 20742}

\author{Zohar Komargodski}
\affiliation{School of Natural Sciences, Institute for Advanced Study,\\
 Princeton, NJ 08540}

\date{\today}

\begin{abstract}
Anomalously small gaugino masses are a common feature of various
models of direct gauge mediation. This problem is closely related
to the vacuum structure of the theory. In this paper we show that
massive SQCD can have SUSY-breaking vacua which are qualitatively
different from the ISS vacuum. These novel vacua are metastable
with respect to decay to the ISS vacuum. We demonstrate the
possibility of addressing the gaugino mass problem in this
framework.  We study the properties of these vacua and construct
an example of a model of direct gauge mediation.
\end{abstract}

\maketitle

\section{Introduction and summary of results}

Gauge mediation of supersymmetry (SUSY) breaking is a compelling
scenario~\cite{Dine:1981gu,Dine:1982zb,AlvarezGaume:1981wy,Nappi:1982hm}.
It addresses the flavor puzzle and often has the virtue of being
calculable. Its distinctive footprints were summarized in the form
of sum rules in~\cite{Meade:2008wd} and further explored
in~\cite{Carpenter:2008wi,Carpenter:2008he,Buican:2008ws,Rajaraman:2009ga}. Direct gauge
mediation~\cite{Affleck:1984xz} is especially appealing since
there is no need to add a separate sector of messengers; they
emerge from the dynamics.

However, we are still far from having satisfactory complete
models. Gauge mediation is generally afflicted by the $\mu/B_\mu$
problem\footnote{This has recently been discussed
in~\cite{Csaki:2008sr,Komargodski:2008ax,Mason:2009iq}.} and by
the Landau pole problem.\footnote{This is ameliorated in theories
with a sector of
messengers~\cite{Dine:1993yw,Dine:1994vc,Dine:1995ag,Seiberg:2008qj,Elvang:2009gk}.}
A somewhat less well-known (but still very common) problem is the
unexpected smallness of the visible gaugino masses relative to the
masses of the scalars. Since the gauginos cannot be lighter than
the electroweak scale the scalars are rendered heavy,
exacerbating the need for fine-tuning.

Recently the problem of gaugino masses has been studied
 in~\cite{Komargodski:2009jf}. It has been shown that the smallness
of gaugino masses is closely related to global properties of the
vacua of the theory. In particular, it has been shown that models
which break SUSY in the lowest energy state of the low-energy
effective theory (namely, the renormalizable theory around the
SUSY-breaking vacuum) necessarily have anomalously small gaugino
masses. More precisely, the contributions to the gaugino masses
vanish at leading order in SUSY breaking.

The problem of gaugino masses is rather pervasive in models of
direct gauge mediation. For example, in models based on massive
SQCD~\cite{Intriligator:2006dd} 
the absence of gaugino masses at
leading order in SUSY breaking (and the resulting split-SUSY-like
spectrum) has been observed in a few examples,
e.g.~\cite{Csaki:2006wi,Abel:2007jx,Haba:2007rj,Essig:2008kz}.\footnote{Other
classes of examples where this phenomenon occurs are, for
instance,~\cite{Izawa:1997gs,Seiberg:2008qj,Elvang:2009gk} and in
a holographic realization of gauge mediation
\cite{Benini:2009ff}.} In light of the discussion above, the
reason is clear. The ISS vacuum is the
ground state of the renormalizable theory at the IR. The existence
of a SUSY vacuum due to non-perturbative effects very far in field
space is not sufficient to generate gaugino masses. Given the
arguments above we are therefore motivated to study different
vacua of massive SQCD; such vacua have a completely different kind
of metastability than the original example of ISS (and many
deformations of it).

To find vacua of this sort we abandon the usual strategy of
looking for SUSY-breaking minima. Instead, the approach we pursue
is to look for states in massive SQCD with a higher vacuum energy
than the ISS vacuum.\footnote{Interestingly,
 in this context~\cite{Kitano:2006xg} provided a non-generic
model with a conventional spectrum of soft scales. Indeed, the
solution of~\cite{Kitano:2006xg} is not a ground state of the IR
theory and it falls to the class of theories we are describing.
Here we emphasize the generality of these models and the
inevitability of such metastable structures. We also argue that
such vacua are easy to find in massive SQCD and are easily
rendered generic.}

In this paper we focus on studying the gaugino mass problem in the
context of massive SQCD. We show that the theory contains
pseudomoduli spaces (i.e. classical flat directions in field space) with a higher vacuum energy than the ISS
vacuum. The problem then reduces to constructing vacua on these
pseudomoduli spaces. This can be done in variety of ways and for
the sake of concreteness and providing an existence proof we focus
on certain deformations of massive SQCD. The result is that we
find new SUSY-breaking vacua. An important property of these vacua
is that they can decay to the ISS vacuum as well as to
supersymmetric vacua. These decays are visible in the
renormalizable approximation around our SUSY-breaking vacuum.
Therefore, this vacuum has the right properties to generate
sizeable gaugino masses. We use this SUSY-breaking vacuum to
exhibit a generic\footnote{By ``generic" we mean that one can add
any operator consistent with the symmetries whose coefficient is
of the same order of magnitude as existing operators, without
spoiling the properties of the metastable vacuum.}
 model of gauge mediation based on massive SQCD.

Of course, it should be mentioned that we do not attempt to
construct a model that solves all the difficulties of gauge
mediation. The goal of this paper is to demonstrate that the
problem of gaugino masses can be addressed by reconsidering the
vacuum structure of SUSY-breaking models and more generally the
strategy of model building. We hope that these general guidelines
will lead to further progress in realizing SUSY breaking and its
transmission to the visible sector.

The paper is organized as follows. In
section~\ref{review} we review the relation between gaugino
masses and the vacuum structure. In
section~\ref{toy} we describe our construction and
comment on some phenomenological issues. Finally, we conclude in
section~\ref{con} where we also discuss open problems. Some
technical details on Wess-Zumino models are relegated to an
appendix.


\section{Gaugino masses in gauge mediation}\label{review}
This section is a review of~\cite{Komargodski:2009jf}.
For references and more details the reader
is advised to consult the original paper. Our discussion here is sufficient for the purposes of this work.

One perplexing property of many models of gauge mediation is that
they predict anomalously small gaugino masses. In other words, if
the gaugino masses are set to be around the expected soft scale,
the scalar masses are at the multi-TeV scale. Such heavy scalar
masses are phenomenologically undesirable.

The key point is that the smallness of gaugino masses has to do
with global properties of the theory. We can understand this in a
somewhat simplified setup. Consider the most general theory of
messengers in the $5\oplus \overline 5$ representation of $
SU(5)$,
\begin{equation}\label{messtheory}
W=\lambda^i_{ j} X\psi_i\bar\psi^{ j}+m^i_{ j}\psi_i\bar\psi^{
j}~,
\end{equation}
where $X$ is a spurion, $\langle X\rangle =x+\theta^2F$, and
$\psi,\bar\psi$ are chiral superfields of messengers in $5\oplus
\overline 5$. This has been first discussed in full generality
in~\cite{Cheung:2007es}. The well known paradigm of minimal gauge
mediation is the special case $m^i_j=0$.

As has been shown in~\cite{Ray:2006wk}, in renormalizable theories
the direction labeled by the bottom component of the spurion $X$
is flat at tree-level.\footnote{One may need to perform a
holomorphic change of variables to make this manifest. We assume
that this has been done and write \eqref{messtheory} without loss
of generality.}  This is related to a general result: in vacua
without tachyons, the scalar in the same chiral multiplet with a
massless Weyl fermion is also massless. Since the fermion in the
$X$ multiplet is the massless Goldstino we see that $x$ must be
massless.

Therefore, when discussing predictions of the
model~\eqref{messtheory} it is natural to discuss them as a
function of $x$, since we do not know a-priori where quantum
effects stabilize it. If supersymmetry breaking is small for every
$x$ (i.e.\  the splittings in the messenger multiplets are small)
we can calculate the gaugino mass via \cite{Cheung:2007es}
\begin{equation}\label{gauginomass}
m_{\lambda}\sim \partial_x\log\det\left(x\lambda^i_{ j}+
m^i_{ j} \right)~.
\end{equation}
We see that if $m_\lambda\neq 0$ then $\det\left(x\lambda^i_{ j}+
m^i_{ j} \right)$ is $x$ dependent. However, since it is a polynomial in $x$,
this implies that there is at least one zero of the polynomial, $x_0$,
\begin{equation}\label{gauginoi}
\det\left(x_0\lambda^i_{ j}+
m^i_{ j} \right)=0~.
\end{equation}
Since $x\lambda^i_{ j}+m^i_{j}$ is the mass matrix of the fermion
messengers,~\eqref{gauginoi} means that there is a massless
fermion messenger at $x_0$. As we have remarked, massless fermions
also lead to massless bosons (or tachyons) and therefore there is
also a massless (or tachyonic) spin-$0$ messenger. With a little
more work it can be shown that following the direction of this
scalar the energy can always be decreased (or the messenger is
decoupled from the theory \eqref{messtheory} and can be therefore
disregarded).

We conclude that in order to get $m_\lambda\neq 0$ there must be
states with lower energy in the system, visible already at the
renormalizable level. Equivalently, in order to get non-vanishing
gaugino masses one is led to look for theories where the
pseudomodulus space spanned by $x$ is \emph{not locally-stable
everywhere}. There must be points on the pseudomodulus space where
some messengers are unstable. (However, the true vacuum lies
elsewhere and is of course stable.) One context in which such
theories have arisen in the past is the inverted hierarchy
mechanism
\cite{Witten:1981kv,Banks:1982mg,Kaplunovsky:1983yx,Dimopoulos:1997ww,Agashe:1998wm}.

One important assumption in our argument was that SUSY breaking is
a small effect. In fact, in all the known examples it turns out
that introducing large splittings inside the multiplets does not
help much. In spite of the fact that now $m_\lambda$ are non-zero,
a sizable hierarchy between the gauginos and the scalars remains.
This is reminiscent of the known (in)dependence of gaugino masses
on SUSY breaking in minimal gauge mediation. The change from small
to large SUSY breaking is very
mild~\cite{Martin:1996zb,Giudice:1998bp}.

In more detail, suppose the leading order contribution to the
gaugino masses vanishes. Then, in many examples, it turns out that
even if SUSY breaking is fine-tuned to be the maximal allowed one,
it is possible to achieve $m_{\lambda}/m_{{\rm sfermion}}\sim
1/10$. However, in most of the parameter space the hierarchy is
much more significant (see e.g.~\cite{Dima}). It would be
interesting to understand this phenomenon more generally.

A central theme of this work is the comparison between gaugino
masses and scalar masses. This can be
quantified~\cite{Cheung:2007es} using the concept of ``effective
number of messengers," which we review here for completeness. We
can parameterize the gaugino mass as
\begin{equation}\label{gaupar}
m_{\lambda_r}=\frac{\alpha_r}{4\pi}\Lambda_{G}~,
\end{equation}
with the label $r=1,2,3$ standing for $U(1),SU(2),SU(3)$, respectively. The
mass of the superpartner $\tilde f$ is given by
\begin{equation}\label{scapar}
m_{\tilde f}^2=2\sum_r C_{\tilde
f}^r\left(\frac{\alpha_r}{4\pi}\right)^2\Lambda_{S}^2~,
\end{equation}
where $C_{\tilde f}^r$ is the quadratic Casimir of $\tilde f$ in
the gauge group corresponding to $r$.
With this we can define the effective number
of messengers to be
\begin{equation}\label{effnumber}
{\cal N}_{eff}=\frac{\Lambda_G^2}{\Lambda_S^2}~.
\end{equation}
In the case of minimal gauge mediation ${\cal N}_{eff}$ coincides
with the actual number of messenger superfields. In a more general
setup it does not have to be an integer. It is possible
 to generalize the notion of effective number of messengers 
to theories that respect just the SM gauge symmetry, 
$SU(3)\times SU(2)\times U(1)$, but we will not need it here.

A useful and simple corollary which will be relevant for our
specific example is the following. Consider a theory with a set of
messengers $\psi, \bar \psi $ which are stable (i.e. massive) on
the entire pseudomodulus space, and some other set of messengers
$\varphi,\bar \varphi$ which are tachyonic at some points on the
pseudomodulus space. Then only $\varphi,\bar \varphi$ can
contribute to the gaugino masses, while generally all of the
messengers contribute to the scalar masses. Thus, the addition of
$\psi, \bar \psi $ does not affect $\Lambda_G$ in \eqref{gaupar}
but increases $\Lambda_S$ in \eqref{scapar}. In this case one
finds that the total effective number of messengers is smaller
than the effective number of $\varphi,\bar \varphi$ pairs.


\section{Constructing a New Solution of Massive SQCD}\label{toy}

\subsection{The ISS Solution and Beyond}
Consider SQCD with an $SU(N_c)$ gauge group and $N_f$
(anti-)fundamental electric quarks ($\bar Q_i$) $Q^i$ where
$i=1...N_f$. We suppress the electric color index. Our interest is
in the regime $N_c<N_f< \frac{3}{2} N_c$ where the theory has a known
weakly coupled dual description in the IR~\cite{Seiberg:1994pq}.
The authors of~\cite{Intriligator:2006dd} considered this theory
deformed by a mass term in the UV,
\begin{equation}\label{mass}
    W=mQ^i\bar Q_i~.
\end{equation}
Denoting by $\Lambda$ the strong coupling scale of the theory, we
have to assume that $m\ll \Lambda$ for the analysis below to be
valid.
 Near the origin of field space, the dynamics of this theory 
at low energies is captured by an $SU(N)$
 gauge theory, $N\equiv N_f-N_c$, with $N_f$   
(anti-\nolinebreak[4]) fundamental magnetic quarks 
($\bar q^i$) $q_i$ where $i=1...N_f$.
 There is also an $N_f\times N_f$ meson field $\Phi_j^i$ transforming as a singlet under the gauge group. The superpotential at low energies is
 \begin{equation}\label{seibergdula}
    W=hq_i\Phi_j^i\bar q^j-h\mu^2\Phi_i^i~,
 \end{equation}
where $\mu^2=-m\Lambda$. The kinetic terms for these emergent
degrees of freedom are canonical.

It has long been known that this theory has no SUSY vacuum at
$\langle\Phi\rangle=0$. This can be seen due to the rank
condition: looking at the $F$-terms of the meson components,
\begin{equation}\label{Fterms}
    F_i^{j}=hq_i\bar q^j-h\mu^2\delta_i^j~,
\end{equation}
we see that the two terms cannot cancel each other due to the fact
that the rank of the first matrix is at most $N_f-N_c$. The
remarkable discovery of~\cite{Intriligator:2006dd} is that in fact
the origin is a SUSY-breaking solution. In this vacuum the
magnetic quarks $q$ obtain expectation values of the order $ \mu$.
This is much smaller than $\Lambda$ and therefore does not mix
with the intricate physics at the strong coupling scale.

Let us use the following notation for the meson field and the magnetic quarks:
\begin{gather}\label{issdecomp}
\Phi=\left( \begin{array}{cc}
(V)_{N\times N} & ( Y)_{N\times (N_f-N)} \\
(\bar Y)_{(N_f-N)\times N} & (Z)_{(N_f-N)\times (N_f-N)}
\end{array} \right)~, \cr
q = \left( \begin{array}{cc} \chi_{N\times N} & \rho_{N\times
(N_f-N)}
\end{array}\right)~,\qquad
\bar q = \left( \begin{array}{cc}
\bar \chi_{N\times N} \\
\bar \rho_{(N_f-N)\times N}
\end{array} \right)~.
\end{gather}
The solution found by ISS is constructed to cancel as many of the
$F$-terms \eqref{Fterms} as possible. Namely, we choose the matrix
$q_i\bar q^j$ to have the maximal possible rank, $N$. It turns out
that by doing so the components of the matrix $Z$ in
\eqref{issdecomp} remain undetermined. In other words, the
solution is given by
\begin{equation}\label{ISSsol}
     \Phi=\begin{pmatrix}0_{N\times N} & 0 \cr  0 & Z_{(N_f-N)\times (N_f-N)}\end{pmatrix}~,\qquad q_i\bar q^j=\begin{pmatrix}\mu^2 \mathbb{I}_{N\times N} & 0\cr 0 & 0_{(N_f-N)\times (N_f-N)} \end{pmatrix}~.
\end{equation}
As we have mentioned in section~\ref{review}, classical solutions
of such theories are always accompanied by at least one complex
flat direction. Indeed, there are a few massless particles of the
theory encountered upon expanding around~\eqref{ISSsol}. The most
important one for us is the $N_c\times N_c$ bottom block of the
meson field, $Z$. The one-loop Coleman-Weinberg potential for all
the massless modes can be calculated and one gets that the
metastable state is located at
\begin{gather}\label{Isssolfinal}
\Phi=0~,\qquad  q = \mu\left( \begin{array}{cc}
\mathbb{I}_{N\times N} & 0_{N\times (N_f-N)}
\end{array}\right)~,\qquad
\bar q = \mu\left( \begin{array}{cc}
\mathbb{I}_{N\times N} \\
0_{(N_f-N)\times N}
\end{array} \right)~.
\end{gather}
This Higgses the magnetic gauge group completely.

For us it will be important to understand some global properties
of the theory \eqref{ISSsol}. At one-loop, the theory around the
pseudomoduli directions $Z_i^j$ can be described as a set of
decoupled O'Raifeartaigh-like models. In the variables of
\eqref{issdecomp} this is given by
\begin{equation}\label{toylag}
    \frac1{h}W=-\mu^2 Z_i^i+Z_i^j\rho_j^c\bar\rho^i_c+\mu \rho_i^c\bar Y^i_c+\mu Y_i^c\bar \rho^i_c~.
\end{equation}
$SU(N_f-N)$ flavor symmetry together with the fact that we
consider only one-loop diagrams guarantees that the result is only
a function of single-trace combinations of the form $\tr\,
((ZZ^\dagger)^n)$ and therefore it is enough to consider only one
component of the matrix $Z_i^j$ together with the fields coupled
to it. For example, consider $Z_1^1$. Then the model we get is
just $N$ copies of the theory:
\begin{equation}\label{toylagi}
   \frac1{h} W=-\mu^2 Z+Z\rho\bar \rho+\mu \rho\bar Y+\mu Y\bar \rho~,
\end{equation}
where we have stripped all the indices for simplicity. As we will
see, to understand many of the interesting features of massive
SQCD it is enough to consider~\eqref{toylagi}.

Setting $\rho=Y=\bar \rho=\bar Y=0$ we see that $Z\in \mathbb {C}$
is a complex line of degenerate classical solutions with vacuum
energy $V=h^2|\mu|^4$. It can be easily seen that, as a function of
$Z$, the particles $\rho,Y,\bar \rho,\bar Y$ are always
non-tachyonic. An equivalent property is that the fermions in
these chiral multiplets are strictly massive for any $Z$. As
explained in~\cite{Komargodski:2009jf} these two properties result
from the complex line $Z$ being a local ground state of the
theory, or in other words, \emph{``a locally stable pseudomodulus
space."} We have reviewed in section~\ref{review} the reason for
why these theories lead to anomalously small values for the
gaugino masses.

Suppose the visible sector gauge group is embedded in the flavor
symmetry group of~\eqref{toylag}. Then, $\rho,\bar \rho, Y, \bar
Y$ transform in some vector-like representation while $\tr\, Z$ is
neutral. Regardless of the value of $\langle \tr\, Z\rangle$ at
which the effective potential stabilizes the modulus, the leading
order contribution to the gaugino mass vanishes. Many deformations
of the basic starting point~\eqref{mass} were constructed and the
dynamics essentially boiled down to shifting the original ISS
vacuum $\langle \tr\, Z\rangle =0$ to some non-zero value of
$\langle \tr\, Z\rangle$. Of course, in all these cases it was
obtained that the gaugino mass vanishes at leading
order~\cite{Csaki:2006wi,Abel:2007jx,Haba:2007rj,Essig:2008kz}.
\footnote{In fact, for $\langle \tr\, Z\rangle=0$ there is a
restoration of an R-symmetry under which $R(Z)=2$ and therefore
the gaugino masses are exactly zero. For nonzero $\langle \tr\,
Z\rangle$ there is no unbroken R-symmetry but the leading
contributions to gaugino masses still vanish.}

We see here very explicitly that to obtain comparable masses for
the scalars and the gauginos, a different strategy has to be
invoked. In the next subsection we describe a different
pseudomoduli space of the theory~\eqref{mass}. Along this
pseudomoduli space the vacuum energy is higher than the vacuum
energy of the ISS solution. In addition, this pseudomoduli space
is not locally stable everywhere. At some regions of the
pseudomoduli space there are tachyons, but the metastable state
could be in a different place. This ``global" property of the
space of classical solutions is essential for obtaining a leading
order contribution to gaugino masses, and indeed we will see that
upon finding a local minimum on our lifted pseudomoduli space,
gaugino masses are generated.

\subsection{A Different Pseudomoduli Space of Massive SQCD}

Consider again the theory defined by~\eqref{seibergdula}.
The solution of ISS is constructed to cancel as many of the
$F$-terms~\eqref{Fterms} as possible. Indeed, the rank of the
quark matrix $q_i\bar q^j$ is at most $N$, which is exactly
saturated by~\eqref{ISSsol}. Now let us consider the case where
the rank of the matrix $q_i\bar q^j$ is
 \begin{equation}\label{newrank}
{\rm rank}(q_i\bar q^j)=k,\qquad  0<k\leq N~.
\end{equation}
For $k=N$ we recover ISS, but here we are interested in $k<N$.
Related ideas were discussed in~\cite{Giveon:2007ef,
Giveon:2007ew,Essig:2008kz}.

From now on it will be convenient to
use the following parametrization for the
meson superfield and the magnetic quarks:
\begin{gather}\label{newdecomp}
\Phi=\left( \begin{array}{cc}
(V)_{k\times k} & ( Y)_{k\times (N_f-k)} \\
(\bar Y)_{(N_f-k)\times k} & (Z)_{(N_f-k)\times (N_f-k)}
\end{array} \right)~, \cr
q = \left( \begin{array}{cc}
(\chi_1)_{k\times k} & (\rho_1)_{k\times (N_f-k)} \\
(\chi_2)_{(N-k)\times k } & (\rho_2)_{(N-k)\times (N_f-k)}
\end{array}\right)~,\qquad
\bar q = \left( \begin{array}{cc}
(\bar \chi_1)_{k\times k} & (\bar \chi_2)_{k\times (N-k)} \\
(\bar \rho_1)_{(N_f-k)\times k} & (\bar \rho_2)_{(N_f-k)\times
(N-k)}
\end{array} \right)~.
\end{gather}
To realize~\eqref{newrank} we look at the following classical
solution:
\begin{gather}\label{newsol}
\Phi=\begin{pmatrix}0_{k\times k} & 0 \cr  0 & Z_{(N_f-k)\times
(N_f-k)}\end{pmatrix}=0~,\cr  q = \mu \left( \begin{array}{cc}
\mathbb{I}_{k\times k} & 0_{k\times (N_f-k)} \\
0_{(N-k)\times k } & 0_{(N-k)\times (N_f-k)}
\end{array}\right)~,\qquad \bar q =
\mu \left( \begin{array}{cc}
\mathbb{I}_{k\times k} & 0_{k\times (N-k)} \\
0_{(N_f-k)\times k} & 0_{(N_f-k)\times (N-k)}
\end{array} \right)~.
\end{gather}
The magnetic gauge group is now only partly Higgsed to $SU(N-k)$.
All the $\chi$ quarks acquire mass either from tree-level terms or
from the super-Higgs mechanism. The massless particles in the
vacuum~\eqref{newsol} are thus only the components of the
$(N_f-k)\times (N_f-k)$ matrix $Z$. The energy of the degenerate
classical solutions labelled by $Z$ is higher than the one of ISS,
\begin{equation}\label{enediff}
    \Delta V=h^2(N-k)|\mu|^4~.
\end{equation}
Near the origin of $Z$, there are some tachyonic magnetic quarks
coming from combinations of $\rho_2$, $\bar\rho_2$. Following
these tachyons we reach the ISS vacuum. However, as we survey the
pseudomoduli space spanned by $Z$ we find that, schematically, for
$Z>\mu$ all the magnetic quarks are massive.
We shall next discuss it in more detail.

Similarly to the analysis around~\eqref{toylag}
and~\eqref{toylagi}, the theory on this pseudomoduli space
effectively factorizes into a sum of generalized O'Raifeartaigh
models. As before, $SU(N_f-k)$ flavor symmetry constrains the
dependence on $Z$ so it is enough to look at a single entry of the
matrix, e.g. $Z_1^1$. Doing so we get the following theory (where
we omit the indices from $Z$, as in~\eqref{toylagi}):
\begin{equation}\label{newOR}
  \frac1{h} W=-\mu^2 Z+\sum_{c=1}^{N-k}Z\rho_2^c(\bar \rho_2)_c +\sum_{c=1}^{k}\left(Z\rho_1^c(\bar \rho_1)_c+\mu  \rho_1^c\bar Y_c+\mu Y^c (\bar\rho_1)_c\right)~.
\end{equation}
Thus, we have $k$ copies of models similar to those
in~\eqref{toylagi}. However, now there are also $N-k$ copies of
sectors with the magnetic quarks $\rho_2,\bar\rho_2$. These
sectors are qualitatively different for reasons we will now
explain. Contrary to the theory in~\eqref{toylag},~\eqref{toylagi},
as mentioned above, here for $Z<\mu$
some combinations of the $\rho_2,\bar \rho_2$ particles are
tachyonic.

Suppose some dynamics stabilizes the pseudomoduli $Z$ farther
away from $\mu$. Then this does give rise to a well defined
metastable state and because of the global properties of the
pseudomoduli space, we do have a leading order contribution to
the gaugino masses. The way to see this is to note that the
$\rho_2,\bar \rho_2$ fields look like the messengers of minimal
gauge mediation (see~\cite{Giudice:1998bp} for review) and would
give the well known $\mu^2/\langle Z\rangle$ contribution to
$m_\lambda$, rendering the scalar and gaugino masses comparable
parameterically.

In general, theories of the form~\eqref{newOR} are of the type
classified and discussed in~\cite{Cheung:2007es}, so this branch
of massive SQCD also gives a dynamical realization for some
particular example of (Extra)-Ordinary Gauge Mediation (EOGM).

The pure massive SQCD theory~\eqref{mass} contains at the tree
level of the low energy description the pseudomoduli
space~\eqref{newsol} as well as the accompanying
structure~\eqref{newOR}, however, the dynamics of the theory does
not naturally give rise to a solution at some permissible value of
$Z$ (where there are no tachyonic modes). Rather, the effective
potential as a function of $Z$ pushes the theory towards the
region with tachyons and eventually to the ISS state.

Thus, as a model building quest, the situation is completely
analogous to the question of breaking the R-symmetry in ISS. The
theory has to be deformed or modified in one out of many possible
ways. However, achieving this goal here will also give rise to a
conventional spectrum of superpartners, so the typical split-SUSY
spectrum encountered so far can be avoided.

The rest of the note is dedicated to a description of one
particular choice of a deformation that allows to stabilize at
permissible values of $Z$.

\subsection{Deforming the Model}

As we have mentioned, the theory~\eqref{mass} on the pseudomoduli
space~\eqref{newsol} with $k<N$ has no metastable SUSY-breaking
states. This conclusion remains true even if~\eqref{mass} is
deformed by some small non-renormalizable operators in the UV
(which may flow to renormalizable operators in the IR with small
coefficients).

Let us explain why this is the case. For $Z<\mu$ there are
tree-level tachyons, hence there are no metastable states in this
regime. On the other hand, the result of Appendix~\ref{WZapp}
implies that for $Z\gg\mu$ there cannot be SUSY-breaking vacua as
well. In fact, the log approximation discussed in
Appendix~\ref{WZapp} becomes predominant very quickly past $\mu$
and therefore these two arguments, combined, explain the absence
of metastable states in this theory. This has been verified
in~\cite{Essig:2008kz} for a particular deformation, but we see
that this is a general result.

These arguments suggest that we should separate the mass scales of the theory. Indeed, there is no reason to have all the mass parameters of~\eqref{mass} degenerate. Suppose there are two mass scales,
$m_1>m_2$, such that $k$ of the $N_f$ electric quarks have mass $m_1$ and the rest $N_f-k$ quarks have mass $m_2$:
\begin{equation}\label{newelesup}
    W=m_1\sum_{i=1}^{k}Q^i\bar Q_i+m_2\sum_{i=k+1}^{N_f}Q^i\bar Q_i~.
\end{equation}
The global non-Abelian flavor symmetry is $SU(k)\times SU(N_f-k)$. As
before, we consider $k<N$ for the analysis below to be valid.

The dual description is:
 \begin{equation}\label{seibergdulnew}
    W=hq_i\Phi_j^i\bar q^j-h\mu_1^2\sum_{i=1}^k\Phi_i^i-h\mu_2^2\sum_{i=k+1}^{N_f}\Phi_i^i~,
 \end{equation}
with $\mu_1^2=-m_1\Lambda$, $\mu_2^2=-m_2\Lambda$.
Of course, this has an $SU(N)$ gauge symmetry
whose indices we suppress.
One may expand around
the pseudomoduli space of classical solutions
\begin{gather}\label{newsol1}
    q_i\bar q^j=\begin{pmatrix}\mu_1^2 \mathbb{I}_{k\times k} & 0\cr 0 & 0_{(N_f-k)\times (N_f-k)} \end{pmatrix}~,\qquad  \Phi=\begin{pmatrix}0_{k\times k} & 0 \cr 0 & Z_{(N_f-k)\times (N_f-k)}\end{pmatrix}=0~.
\end{gather}
The gauge symmetry is Higgsed to $SU(N-k)$. The surviving non-Abelian
global symmetry is $SU(k)_d\times SU(N_f-k)$, where
$SU(k)_d$ is a diagonal mixture of the original
$SU(k)$ and some broken gauge generators.

Studying the fluctuations around~\eqref{newsol1} we find a
structure very similar to~\eqref{newOR}, but now the mass
scales are not all the same:
\begin{equation}\label{newORi}
    \frac1{h}W=-\mu_2^2 Z+\sum_{c=1}^{N-k}Z\rho_2^c(\bar \rho_2)_c +\sum_{c=1}^{k}\left(Z\rho_1^c(\bar \rho_1)_c+\mu_1  \rho_1^c\bar Y_c+\mu_1 Y^c (\bar\rho_1)_c\right)~.
\end{equation}
The pseudomodulus space is again characterized by $Z$, with
vanishing expectation values for all the other fields.

The theorem of Appendix~\ref{WZapp} is still true, regardless of
how the theory is deformed: there cannot be a metastable vacuum
for $Z$ bigger than all the mass scales of the theory, namely,
$Z>\mu_1$. However, now there are tachyons only when $Z<\mu_2$.
(Some combinations of the fields $\rho_2,\bar \rho_2$ become
tachyonic in this region.) We therefore have the nontrivial region
\begin{equation}\label{regime}
    \mu_2<Z<\mu_1~,
\end{equation}
where there are no tachyons and there is no
argument for the absence of metastable solutions. Indeed, we will
see that it is possible to deform the theory by small
non-renormalizable operators in the UV and obtain SUSY-breaking
solutions in this interval.

As we have mentioned, the problem at this stage is very closely
related to the question of how to break the R-symmetry in the
original ISS solution. Our choice for the solution of the problem
(which is by no means unique) is reminiscent of the work
of~\cite{Essig:2008kz} in the context of R-symmetry breaking in
the ISS vacuum.

We deform the UV theory~\eqref{newelesup} by gauge invariant
operators quartic in the electric quarks. For concreteness we
consider the operator
\begin{equation}\label{UVop}\delta W=\frac{1}{M_*} \left(
\sum_{i=k+1}^{N_f} Q^{i} \bar Q_{i} \right)^2~,
\end{equation}
where $M_*$ is some high energy scale. In the IR this becomes
\begin{equation}\label{IRop}
\delta W=\frac{\epsilon h\mu_2}{2}(\tr\, Z)^2~,\qquad \epsilon\sim
\frac{\Lambda^2}{M_*\mu_2}~.
\end{equation}
With this deformation we are led to consider the theory
\begin{equation}\label{newORoii}
   \frac1{h} W=-\mu^2_2 Z+\sum_{c=1}^{N-k}Z\rho_2^c(\bar \rho_2)_c +\sum_{c=1}^{k}\left(Z\rho_1^c(\bar \rho_1)_c+\mu_1
\rho_1^c\bar Y_c+\mu_1 Y^c (\bar\rho_1)_c\right)+
\frac{\epsilon\mu_2}{2}Z^2~.
\end{equation}
We will see that $\epsilon$ is a parameterically small number and therefore the scale $\epsilon\mu_2$ is much smaller than all the other low
energy scales in the problem.\footnote{Replacing $(\tr\, Z)^2$ in
\eqref{IRop} by $Z^2$ in~\eqref{newORoii} is not quite precise,
but it is done for simplicity. One can analyze the complete system
and arrive at similar conclusions.} Such deformations may not leave any R-symmetry in the problem, but as we have explained, the difficulty of getting gaugino masses is not related to R-symmetry. 

\subsection{The Dynamics of the Deformed Model}

The basic dynamics of~\eqref{newORoii} can be understood from
general considerations which we outline here. These considerations
are enough to establish the existence of metastable states in the
theory~\eqref{newORoii} and in turn in the complete theory.

Without the deformation $\delta W=\frac{\epsilon h\mu_2}{2}Z^2$, the
$Z$ field is flat at tree level. Since $\epsilon$ is assumed to be
small, the only relevant effect of this deformation at tree level
is to introduce a tadpole for $Z$:
\begin{equation}\label{tadpole}
    V_{{\rm tree}}=-\epsilon h^2\mu_2^3 Z+c.c.~.
\end{equation}
The basic idea is to balance this tree-level tadpole, which is
parameterically suppressed by $\epsilon$, with the one-loop
effects, which are $\epsilon$ independent. There are two different
types of one-loop contributions. One is from the fields
$\rho_1,\bar \rho_1$ and $Y,\bar Y$ which are coupled to each
other. This is denoted by $V^{(1)}_{{\rm one-loop}}(Z)$. The other
is from the fields $\rho_2,\bar \rho_2$, which we denote
$V^{(2)}_{{\rm one-loop}}(Z)$.  The total scalar potential is thus
\begin{equation}\label{totalpot}
    V(Z)=V_{{\rm tree}}(Z)+V^{(1)}_{{\rm one-loop}}(Z)+V^{(2)}_{{\rm one-loop}}(Z)~.
\end{equation}
The function $V^{(2)}_{{\rm one-loop}}(Z)$ does not exist for
$Z<\mu_2$ due to the fact that some components of $\rho_2,\bar
\rho_2$ become tachyonic and cannot be integrated out. We
therefore consider the potential $V(Z)$ only in the regime
$Z>\mu_2$. In this regime, $V^{(2)}_{{\rm one-loop}}(Z)$ can be
crudely approximated by the leading log. (This is rigorously
justified if $Z\gg\mu_2$.)

Similarly, in the regime~\eqref{regime},
$V^{(1)}_{{\rm one-loop}}(Z)$ is roughly approximated by a
quadratic function. In fact, the quadratic approximation is
formally valid in the limit $Z\ll\mu_1$, but our purpose here is
to obtain a gross understanding of the physics so this is good
enough. We therefore get that the potential we should study is
\begin{equation}\label{totalpoti}
    \frac1{h^2}V(Z)=-\epsilon \mu_2^3 Z-\epsilon \mu_2^3 Z^\dagger+(N-k)\frac{\alpha_h}{4\pi}\mu_2^4\log(|Z|^2)+k\frac{\alpha_h}{4\pi}\frac{\mu_2^4}{\mu_1^2}|Z|^2~,
\end{equation}
where $\alpha_h=\frac{h^2}{4\pi}$. We do not include here various
order one numerical coefficients from the one-loop potential in
order not to clutter the expressions, however, all the signs are
correct.

The potential~\eqref{totalpoti} has a minimum if~\footnote{For simplicity,
we take all the parameters to be real and positive.}
\begin{equation}\label{ineq}
    \frac{N-k}{k} \lesssim \left(\frac{Z}{\mu_1}\right)^2~,
\end{equation}
and if
\begin{equation}\label{epsvalue}
   \sqrt{k(N-k)}\frac{\alpha_h}{4\pi}\frac{\mu_2}{\mu_1} \lesssim\epsilon\lesssim k\frac{\alpha_h}{4\pi}\frac{\mu_2}{\mu_1}~.
\end{equation}
Since $Z/\mu_1<1$ for our approximation to make qualitative sense,
from \eqref{ineq} we see that we need $k>N-k$, which means that we
should expect to have more of the $\rho_1-Y$-type sectors than the
$\rho_2$-type fields. From~\eqref{epsvalue} we see that,
self-consistently with our assumption, $\epsilon$ is
parameterically smaller than any other tree-level quantity in the
Lagrangian and, therefore, the backreaction from the quadratic
deformation is indeed negligible. The value of $\epsilon$ is
almost determined in terms of $h$ and $\mu_2/\mu_1$; if $\epsilon$
is not in the range~\eqref{epsvalue}, then either there is no
minimum or it is outside of the range of validity of our
approximation. (An exact treatment shows that there is typically
no minimum in this case.)

Of course, the model~\eqref{newORoii} can also be solved exactly
at one-loop and it has been verified that all its qualitative
features agree remarkably well with the analysis above. The
theory~\eqref{newORoii} also contains solutions with lower vacuum
energies, exactly like the parent SQCD theory. In one of them the
$\rho_2,\bar \rho_2$ fields obtain nonzero VEVs. This corresponds
to the ISS solution. The other is where
$Z=\frac{\mu_2}{\epsilon}$, where we have a new SUSY vacuum. In
both cases the separation in field space is at least of order
$\mu_1$ and the energy difference is $\mu_2$, therefore the
parametric separation of scales guarantees longevity.

\subsection{Direct Mediation of SUSY Breaking}

Imagine that we embed the $SU(5)$ GUT group in the flavor symmetry
group $SU(N_f-k)$. Then, some components of $ \rho_1, \rho_2, Y$
and $\bar \rho_1, \bar \rho_2, \bar Y$ are in the $5$ and $\bar 5$
of $SU(5)$, respectively. The matrix $Z$ decomposes to a matrix in
the adjoint, $5$, $\bar 5$ and the singlet representation of
$SU(5)$. Let us study the branch where $\tr\, Z\neq 0$ and the GUT
group remains unbroken.

We can then calculate the scalar and gaugino masses of the
model~\eqref{newORi}. The various components of the matrix $Z$ do
not have a supersymmetric spectrum but their contributions to the
soft terms in the MSSM are negligible. The reason is that the
typical mass of the scalars in $Z$, hence the typical mass
splitting in the supermultiplet, is
$\sqrt{\frac{\alpha_h}{4\pi}}\frac{\mu_2^2}{\mu_1}$. This has a
negligible effect on the visible soft masses.

Therefore, for our purposes we may forget about the matrix $Z$.
The messengers are only the magnetic quarks $\rho_{1,2}$, their
conjugates $\bar\rho_{1,2}$ and the meson components $Y$, $\bar
Y$. These types of models have been analyzed in EOGM and formulae
for the gaugino and scalar masses in the regime $Z\gg\mu_2$ were
presented (see equations~(2.5),(2.6) in~\cite{Cheung:2007es}). We
can parameterize the ratio of gaugino to scalar masses with the
effective number of messengers ${\cal N}_{eff}$, defined
in~\eqref{effnumber}. The result is:
\begin{equation}\label{effmess}
    \left({\cal N}_{eff}\right)^{-1}=\frac{|z|^2}{(N-k)^2}
\left(\frac{N-k}{|z|^2}+\frac{2k}{|z|^2+4}+
\frac{2k\log \left(\frac{|z|^2+2+
\sqrt{|z|^4+4|z|^2}}{|z|^2+2-\sqrt{|z|^4+4|z|^2}}
\right)}{\left(|z|^2+4\right)\sqrt{|z|^4+4|z|^2}}\right)~,
\end{equation}
where we denote $z=Z/\mu_1$. Note that the $\rho_1-Y$-type
messengers do not generate gaugino masses at leading order (which
is in accord with the discussion in section~\ref{review}) but do
contribute to the scalar masses.  This is the reason for the
inequality ${\cal N}_{eff}\leq N-k$. We therefore end up with
theories that have no parametric suppression of gaugino masses,
but the effective number of messengers is bounded by $N-k$.

\subsection{Comments on Phenomenology}

These theories generically suffer from the Landau pole problem.
For example, if $k=\nolinebreak[4]N-\nolinebreak[4]1=7$, one finds
$15$ messengers and all the representations coming from the matrix
$Z$. The latter can be lifted by introducing another deformation
$\delta W=\epsilon_{ad}h\mu_2\tr\left(Z^2\right)$, which is
linearly independent of $(\tr\, Z)^2$.  This can be arranged so
that all the components of the matrix $Z$ but the trace get in the
IR a large enough mass. In this way, all the contributions to the
beta functions from the matrix $Z$ can be pushed to higher
energies. Consequently, for appropriately chosen $\epsilon_{ad}$,
the Landau pole can be pushed to lie well above the strong coupling scale $\Lambda$.

For $k<N-1$, the Landau pole problem is more
severe. This is simply because the overall number of messengers is
scaled with the number of unbroken color generators, which is
$N-k$. One could perhaps try to address the Landau pole problem as
in~\cite{Abel:2008tx}.

Let us discuss some scales in the problem. Hereafter we assume that
$h\sim 1$. First, the gaugino mass
scale is given by
\beq m_{\lambda_r} \sim \frac{\alpha_r}{4\pi}
\frac{\mu_2^2}{\mu_1} \sim 100\ {\rm GeV}\ \
\ \Rightarrow \ \ \
\frac{\mu_2^2}{\mu_1}\sim 10^5\ {\rm GeV}~.
\eeq
In order to get
$\epsilon$ to be of the right order of magnitude, and if we choose
$M_*\sim \mpl$, we get:
\beq
\epsilon \mu_2 \sim
\frac{h^2}{16\pi^2} \frac{\mu_2^2}{\mu_1} \sim
\frac{\Lambda^2}{\mpl} \ \Rightarrow \  \Lambda\sim 10^{11}\
{\rm GeV}~.
\eeq
Now let us consider the constraints from
calculability. A typical value of the incalculable contributions
to all the masses from corrections to the K\"ahler potential is
$\delta m^2\sim\frac{\mu_2^4}{\Lambda^2}$. Comparing with the
one-loop contributions, which are of the order
$\frac{h^2}{16\pi^2} \frac{\mu_2^4}{\mu_1^2}$, we see that we
should demand $h\Lambda \gg 4\pi \mu_1$, and consequently
$\mu_1\apprle 10^9\ {\rm GeV}$.

Indeed, we can take $\mu_1\sim 10^9\ {\rm GeV}$ and, therefore,
$\mu_2\sim 10^7\ {\rm GeV}$. In this case, by choosing an
appropriate $\epsilon_{ad}$, the Landau pole can be pushed above
$\Lambda$. Relaxing the condition
$M_*\sim \mpl$ one can do better by considering more general
constructions. As a numerical example, for $k= 7$ and $N=8$ we
find metastable states (by choosing an appropriate value for
$\epsilon$, as in~\eqref{epsvalue}). Calculating the value of the
pseudomodulus $\tr\, Z$ at the vacuum  and using the
formula~\eqref{effmess}, we can typically obtain ${\cal N}_{eff}\sim
1/4$. Note that this means that the actual hierarchy in masses is
just a factor $\sim 1/2$ on top of the usual prediction of minimal
gauge mediation with one messenger. This is better than what can
be obtained from studying metastable states on locally stable
pseudomoduli spaces.

In spite of the fact that we analyzed a particular set of quartic
operators in the UV, it is possible to see that the dynamics of
the model is only slightly affected by other small perturbations
consistent with the symmetries. In this sense, the mechanism above
is generic.

Of course, as many of the model building attempts based on on ISS,
this has a few unappealing features. One is the difficulty in
attaining perturbative coupling unification. Second, like in many
other examples, we need to deform the model and tie the
deformation parameters with other low energy scales (which seems
unnatural and unappealing). Last but not least, in the simplest
form of such models, the mass scale for the electric quarks which
one needs to introduce in the UV spoils the main motivation for
dynamical SUSY breaking (which is to explain the smallness of the
electroweak scale).

Here we did not attempt to solve these problems but rather to
demonstrate a new principle in model building, which has shown to
be fruitful in surmounting one pervasive problem of gauge
mediation: the anomalously small gaugino masses. This problem is
interesting and perhaps special as it is connected to the
structure of the vacua in the theory.


\section{Conclusions and Outlook}\label{con}

In this paper we have discussed pseudomoduli spaces of metastable
SUSY-breaking solutions of massive SQCD. The main purpose has been
to establish an existence proof for such vacua and to demonstrate
their crucial phenomenological difference from the more
conventional approach. These vacua are metastable already in the
renormalizable approximation and lead to sizeable gaugino masses.

This is to be contrasted with the ISS solution (and deformations
thereof) which are generally afflicted by anomalously small
contributions to the gaugino masses. We have repeated the argument
of~\cite{Komargodski:2009jf} and have explained why this is the
case. We identified different possible states of massive SQCD
which are metastable even within the low energy effective theory
(however, longevity can be ensured) and solve the parametric
suppression of the gaugino masses. The construction we presented
provides a realization of a particular model of the EOGM type. It
would be nice to understand whether more general models of
messengers can be embedded in dynamical SUSY breaking.

Our discussion on how to find these solutions is general but we choose one particular method to actually demonstrate them.
Another idea which can be implemented in such scenarios was studied recently in \cite{Giveon:2009ur}.
It would be nice to perform a more exhaustive analysis in order to find the most appealing possibility.

Since obtaining sizeable gaugino masses requires having lower
energy states already in the renormalizable approximation (at
least in calculable examples), it would be interesting to study
the cosmological applications of it and the thermal history of
such theories, analogous to the studies
in~\cite{Abel:2006cr,Craig:2006kx,Fischler:2006xh}. This question is clearly very general and deserves a separate study.


\vskip 0.2in \noindent {\bf Acknowledgements:}\ \ We are grateful
to Z.~Chacko, H.~Elvang, D.~Kutasov, D.~Malyshev, N.~Seiberg,
D.~Shih, M.~Strassler, B.~Wecht for insightful comments.  The work
of AG is supported in part by the BSF -- American-Israel
Bi-National Science Foundation, by a center of excellence
supported by the Israel Science Foundation (grant number 1468/06),
DIP grant H.52, and the Einstein Center at the Hebrew University.
AG thanks the CERN Theory Institute and the EFI at the University
of Chicago for hospitality during the course of this work. The
work of AK is partially supported by NSF under grant PHY-0801323.
The work of ZK is supported in part by NSF grant PHY-0503584. Any
opinions, findings, and conclusions or recommendations expressed
in this material are those of the author(s) and do not necessarily
reflect the views of the funding agencies.

\appendix
\section{Wess-Zumino Models at Large Fields}\label{WZapp}

In this appendix we discuss a useful property of Wess-Zumino models at large fields. The simple result we present here is a useful guideline in model building and applies directly to our construction.

Consider a chiral superfield $\phi$ with some superpotential
\begin{equation}\label{supapp}
\int d^2\theta\left( f\phi+\epsilon W(\phi)\right)~,
\end{equation}
where $\epsilon$ is some small parameter and the K\"ahler potential is canonical. A setup of this type is often encountered at low energies in some dual description. E.g. irrelevant operators in the UV can become renormalizable at low energies but they are suppressed by small parameters. The classical potential is
\begin{equation}V=\biggl|f+\epsilon\frac{\partial W}{\partial\phi}\biggr|^2~.\end{equation}
Since an absolute value of a holomorphic function cannot have a
local minimum with non-vanishing value for the function, there
cannot be proper SUSY-breaking minima in \eqref{supapp}. One could
try to couple the field $\phi$ to some chiral superfields $P,\bar
P$ such that the superpotential  becomes
\begin{equation}\label{supapp1}
\int d^2\theta\left( f\phi+h\phi P\bar P+\epsilon W(\phi)\right)~.
\end{equation}
Now, for large $\phi$ the $P,\bar P$ fields are integrated out and generate a nontrivial K\"ahler metric at one loop. For $\phi\gg \sqrt f$ the contribution of the $P,\bar P$ fields is to modify the K\"ahler potential to
\begin{equation}\label{newkahfun}
K=\left(1+c\frac{\alpha_h}
{4\pi}\log(\phi\phi^\dagger)\right)\phi\phi^\dagger~,
\end{equation}
where $c$ is a number which can be extracted by matching with the anomalous dimension (for a more general treatment of such systems
see~\cite{Intriligator:2008fe})
and $\alpha_h=\frac{h^2}{4\pi}$.
This means that now the scalar potential is of the form
\begin{equation}
V=\left(1+c\frac{\alpha_h}{4\pi}
\log(\phi\phi^\dagger)\right)^{-1}\biggl|f+\epsilon\frac{\partial W}{\partial\phi}\biggr|^2~,
\end{equation}
where we have dropped some unimportant constant correction in the prefactor. At leading order in the loop expansion parameter this can also be written as
\begin{equation}\label{final}
V=\left(1-c\frac{\alpha_h}{4\pi}\log(\phi)-
c\frac{\alpha_h}{4\pi}
\log(\phi^\dagger)\right)\biggl|f+
\epsilon\frac{\partial W}{\partial\phi}\biggr|^2
=\biggl|\left(1-
c\frac{\alpha_h}{4\pi}\log(\phi)\right)
\left(f+\epsilon\frac{\partial W}{\partial\phi}\right)\biggr|^2~.
\end{equation}
We see that again this is an absolute value of a holomorphic
function and therefore has no SUSY-breaking minima (i.e. any
classical solution will have  either tachyons or flat directions).
We could have also got this result by performing a holomorphic
change of variables in~\eqref{newkahfun}.

In spite of our discussion above being in the simplest possible
setup, this result is a very general property of sigma models with
canonical K\"ahler potential: when the VEV of some fields is above
all the mass parameters in the Lagrangian it is generally
impossible to find SUSY-breaking minima at one-loop. This is of
course consistent with the known examples. (At two-loops this can
be circumvented,
e.g.~\cite{Giveon:2008wp,Giveon:2008ne,Kitano:2008tm,Amariti:2008uz}.)

\bibliography{dgmlit}
\bibliographystyle{apsper}
\end{document}